%% file: main.tex
\documentclass[sigconf,nonacm]{acmart}
\usepackage{listings}
\usepackage{tabularx} 
\usepackage{makecell}

\definecolor{BG}{RGB}{240,240,240}

\lstdefinestyle{pythonstyle}{
    language=Python,
    backgroundcolor=\color{BG},
    basicstyle=\ttfamily\scriptsize,
    commentstyle=\ttfamily,
    keywordstyle=\ttfamily\bfseries,
    stringstyle=\ttfamily,
    showstringspaces=false,
    breaklines=true,
    breakatwhitespace=false,
    columns=fullflexible,
    keepspaces=true,
    frame=single,
    framerule=0.3pt,
    framesep=4pt,
    rulecolor=\color{black},
    numbers=none,
    tabsize=4,
    captionpos=b,
    aboveskip=0.8em,
    belowskip=0.3em
}

\AtBeginDocument{%
  }

\setcopyright{none}

\begin{document}

\title[Evaluation Bias and Epistemic Inequality in Global Software Development]{Evaluation Bias and Epistemic Inequality in Global Software Development}

\author{Sam Khosravi}
\affiliation{%
  \institution{KTH Royal Institute of Technology}
  \city{Stockholm}
  \country{Sweden}}
\email{samkh@kth.se}

\author{Amir H. Payberah}
\affiliation{%
  \institution{KTH Royal Institute of Technology}
  \city{Stockholm}
  \country{Sweden}}
\email{payberah@kth.se}

\renewcommand{\shortauthors}{Khosravi and Payberah}

\input{sections/0_abstract}




\maketitle
\input{sections/1_introduction}
\input{sections/2_background}
\input{sections/3_method}

\input{sections/4_results}

\input{sections/5_discussion}
\input{sections/6_conclusion}

\section*{Generative AI Usage Statement}
We used generative AI tools to help polish the text and improve clarity.

\section*{Acknowledgment}
 We thank InspireLab at KTH Royal Institute of Technology for partially funding this project.


\bibliographystyle{ACM-Reference-Format}
\bibliography{main}

\appendix

\end{document}

%% file: sections/0_abstract.tex
\begin{abstract}
This paper examines fairness and accountability in global software development by focusing on how competence is assessed and valued across unequal regional contexts. We compare software engineers from East Africa (Rwanda and Uganda) and Northwestern Europe (Sweden and the Netherlands), regions that are increasingly connected but embedded in asymmetric technological, economic, and institutional structures. Despite the rapid growth of African technology ecosystems, empirical evidence on everyday engineering practice and evaluation in these contexts remains limited. We present findings from an on-site mixed-methods study with 48 software engineers across four countries. The study combines programming, system design, and code review tasks with semi-structured interviews. Our results reveal consistent gaps between measured performance and perceived competence. Senior engineers in Rwanda often performed at a level comparable to that of their European peers, yet European participants systematically underestimated the competence of East African engineers. We also observed differences in communication styles and organizational practices across regions, reflecting distinct but complementary ways of working.
\end{abstract}

%% file: sections/1_introduction.tex
\section{Introduction}
Global software development increasingly relies on collaboration across regions with unequal access to resources, infrastructure, and professional recognition. In these settings, assessments of competence shape who is trusted, whose contributions carry weight, and who gains access to career opportunities. Engineers in historically marginalized regions are often evaluated through stereotypes and deficit narratives that frame their work as lower quality or higher risk. These evaluations are not merely individual misunderstandings; they constitute a fairness and accountability problem in the global computing labor market. By shaping whose knowledge is treated as credible and whose expertise is discounted, they contribute to epistemic injustice and procedural unfairness in everyday sociotechnical work practices.

East Africa illustrates this tension clearly. Countries such as Rwanda and Uganda have invested heavily in digital skills and software engineering capacity and are increasingly positioned as sites of global technology work, while Northwestern European ecosystems such as Sweden and the Netherlands remain comparatively well-resourced and institutionally entrenched. Studying these regions together highlights how competence is assessed and compared across unequal contexts. In cross-regional collaboration, perceptions of skill can serve as gatekeeping mechanisms that shape whose work is trusted and whose careers advance, allowing inequitable distributions of credibility and opportunity to persist even when technical evidence suggests overlapping capability.

Despite the relevance of these dynamics, prior work on global software engineering has focused largely on Asia, Eastern Europe, and South America, leaving African contexts underexplored~\cite{Marinho2018,krishna2004managing,merino2017offshoring,iqbal2013outsourcing}. Existing studies on Africa concentrate mainly on South Africa and Kenya~\cite{tanner2009communication,tanner2010software,salami2024software,govender2024african}, while countries such as Rwanda and Uganda remain empirically underrepresented despite major digital infrastructure initiatives~\cite{taarifa2023million}. Moreover, much of this literature relies on secondary data, limiting insight into how perceptions and organizational narratives relate to observed engineering practice~\cite{ajayi2016}. As a result, an important dimension of fairness in computing remains insufficiently examined: how human and organizational evaluation systems distribute trust, credibility, and opportunity across unequal global labor contexts.

To address this gap, we study software engineers in Rwanda, Uganda, Sweden, and the Netherlands to examine how technical performance, communication practices, and competence perceptions interact across unequal contexts. Our analysis is guided by the following research questions:
\begin{itemize}
  \item RQ1: How do software engineers in East Africa and Northwestern Europe differ in coding skills and problem-solving strategies?
  \item RQ2: How do communication styles and organizational practices differ between regions, and how are these differences interpreted in cross-regional collaboration?
  \item RQ3: What forms of perception bias and procedural unfairness arise in cross-regional software development, and how do these biases affect the recognition of competence and the distribution of trust?
\end{itemize}

We adopt a pragmatic mixed-methods approach combining controlled technical assessments with qualitative inquiry. We conducted on-site sessions with 48 software engineers (10-15 per country), who completed a programming task, a system design challenge, and a code review exercise. Programming solutions were analyzed using CodeBERT embeddings~\cite{feng2020codebert}, PCA~\cite{abdi2010principal}, and K-means clustering~\cite{sinaga2020unsupervised}, complemented by manual inspection, while review-comment sentiment was assessed using VADER~\cite{hutto2014vader}. Semi-structured interviews (60-120 minutes) captured developers’ self-assessments, perceptions of peers, and experiences of cross-regional collaboration.

Our findings reveal systematic gaps between measured performance and perceived competence. Senior Rwandan engineers often performed on par with European peers, while junior performance varied more widely. Communication styles differed consistently: East African engineers tended to provide more supportive feedback, whereas European engineers favored more direct critique. Organizational emphases also diverged, with European participants prioritizing upfront design and East African participants emphasizing review and deployment. Crucially, European engineers consistently underrated East African capabilities despite evidence of overlapping competencies, indicating that evaluation practices in global software development can reproduce epistemic and procedural unfairness even when technical contributions warrant recognition.\\

\noindent\textbf{Contributions.} This paper contributes:
\begin{itemize}
  \item Empirical evidence from underrepresented contexts (Rwanda and Uganda) on how competence, credibility, and professional recognition are constructed in global software development.
  \item A comparative mixed-methods analysis linking technical assessments, communication practices, and organizational priorities across East Africa and Northwestern Europe.
  \item Documentation of perception bias and procedural unfairness, showing systematic underrating of East African engineers by European peers despite demonstrated overlap in capability.
  \item Implications for more accountable global collaboration, including how evaluation and feedback practices might be redesigned to reduce bias and better recognize complementary strengths.
\end{itemize}

The remainder of the paper reviews related work, presents the study design and methods, reports results, and discusses implications and limitations.

%% file: sections/2_background.tex
\section{Background and Related Work}
This section provides the theoretical and methodological context for our study. It begins by discussing the evolution of global software engineering and the role of cross-cultural collaboration, followed by an overview of analytical techniques and methodologies relevant to evaluating software engineering skills and practices.

\subsection{Global Software Engineering and Cross-Cultural Collaboration}
Global software engineering has evolved from co-located, in-house teams to geographically distributed collaborations that span continents. Advances in Information and Communication Technology (ICT), cloud-based tools, and agile practices have enabled organizations to recruit talent worldwide and coordinate work across time zones~\cite{ebert2016global,deshpande2016remote,drahokoupil2015outsourcing,oconchuir2009global}. Economic drivers, such as rising labor costs in established markets and the pursuit of cost efficiency and specialized expertise, have further incentivized outsourcing and offshoring~\cite{ebert2016global2}. 

While distributed collaboration offers these benefits, it also introduces challenges of distance, typically categorized as {\em geographical}, {\em temporal}, and {\em socio-cultural}~\cite{saleem2025gsd}. Geographical distance reduces informal communication and shared awareness; temporal distance limits synchronous interaction and delays feedback; and socio-cultural distance introduces differences in work practices, hierarchies, and communication styles. Together, these factors can weaken trust, complicate coordination, and reduce productivity~\cite{stahl2010unraveling, hunley2018impact}. Although agile methodologies and digital tools mitigate some barriers, temporal and cultural differences remain persistent challenges~\cite{zada2015scrum, khan2021agile}.

Cultural variation plays a central role in shaping collaboration in global software development. Hofstede’s cultural dimensions~\cite{macgregor2005impact, hofstede2011dimensionalizing} highlight contrasts in power distance, uncertainty avoidance, and individualism vs. collectivism. For example, high-context communication common in collectivist societies can clash with low-context communication in individualist cultures, often leading to misunderstandings~\cite{Palokangas2013, stahl2010unraveling, mangundjaya2023power}. Studies also show that African and European teams differ in communication and organizational practices, with African teams more often operating within hierarchical structures and European teams favoring flatter ones~\cite{mine_rehabilitation2023}. 

These differences are often compounded by biases, which shape how competence and contributions are perceived and can reinforce inequalities in distributed teams~\cite{Likhi2022, Matthiesen2023, welsch2023navigating}. However, interventions such as cultural competence training and support for multilingual collaboration can help mitigate these barriers~\cite{oluleye2025cross, Wang2024}. Despite this need, most empirical studies on cross-cultural software engineering and outsourcing concentrate on Asia, Eastern Europe, and South America~\cite{sahay2003global, krishna2004managing, iqbal2013outsourcing, alsahli2017agile}. A systematic review of 91 outsourcing papers across 51 journals found no Africa-based studies~\cite{hanafizadeh2020systematic}, and the limited research on Africa largely examines South Africa, Kenya, Nigeria, or North Africa~\cite{tanner2009communication, tanner2010software, salami2024software, govender2024african}. 

Against this backdrop, Rwanda and Uganda are emerging as notable technology hubs in East Africa, driven by large-scale ICT investments, broadband expansion, and supportive policies~\cite{Maresova2016, Pollio2022, Kopati2021, TechGyant2023, MoICTUganda2024}. Rwanda, in particular, has positioned ICT as a national growth engine, with the sector contributing to a 35\% increase in Gross Domestic Product (GDP) in 2023~\cite{TechGyant2023}. Entrepreneurial ecosystems, such as Kigali’s Norrsken hub, further signal international attention to the region’s potential~\cite{Bainomugisha2021, Otioma2019, Mwangi2015, norrsken2025kigali}. Yet empirical research on software engineering practices in these countries remains scarce, leaving Rwanda and Uganda largely absent from comparative global software engineering studies. This underrepresentation motivates our study to generate new insights into how cultural and organizational contexts shape software practices in underexplored regions.

\subsection{Analytical Techniques and Methodologies in Software Engineering Research}

Software engineering research employs a wide range of methods to evaluate both technical competencies and collaborative practices. Quantitative assessments such as standardized coding tests, bug detection tasks, and system design exercises provide objective measures of programming proficiency, algorithmic thinking, and problem-solving strategies~\cite{devlin2010assessing,sahakyan2023general,Garousi2019,Florea2023}. Qualitative approaches, including in-depth interviews and field studies, complement these metrics by capturing organizational norms, cultural dynamics, and perceptions of collaboration~\cite{mclaughlin2001using,lenberg2024qualitative}. Mixed-methods designs are often recommended for their ability to balance objective measurement with contextual richness~\cite{DiPenta2017,betzner2008pragmatic}.

Building on these foundations, recent advances in machine learning and natural language processing have significantly expanded the analytical toolkit for studying software practices at scale. Transformer based models such as CodeBERT~\cite{feng2020codebert}, GraphCodeBERT~\cite{guo2020graphcodebert}, CoCluBERT~\cite{hagglund2021coclubert}, PLBART~\cite{ahmad2021unified}, and CodeT5/CodeT5+~\cite{wang2021codet5,wang2023codet5+} generate code embeddings that enable embedding, retrieval, and clustering of programming solutions and styles. Dimensionality reduction techniques, such as PCA~\cite{abdi2010principal}, further support the discovery and visualization of coding patterns in high-dimensional spaces~\cite{martinez2024graphcodebert}. For collaborative practices, sentiment and affective computing approaches have been applied to pull requests, issue discussions, and code reviews; for example, classifiers such as Senti4SD~\cite{calefato2018sentiment} and SentiCR~\cite{ahmed2017senticr}, along with lexicon-based tools like VADER~\cite{hutto2014vader,obaidi2021development}, have been used to examine tone, supportiveness, and directness in developer communication. Together, these methods enable a holistic assessment of software engineering, capturing technical performance, design reasoning, communication styles, and cultural perceptions.

%% file: sections/3_method.tex
\section{Method}
This study adopts a pragmatic mixed-methods approach~\cite{storey2024guiding}, combining quantitative coding assessments with qualitative interviews to compare software engineers in East Africa (Rwanda and Uganda) and Northwestern Europe (Sweden and the Netherlands). Pragmatism enables the balancing of measurable outcomes with contextual insights, avoiding the rigidity of purely quantitative or qualitative paradigms~\cite{clarke2019pragmatic, feyerabend2020against, law2004after}.

\subsection{Research Questions}\label{sec:rq}
The study is guided by three research questions (RQ), which are formulated based on gaps identified in the literature regarding global software engineering and cross-cultural collaboration with particular attention to the limited representation of East Africa contexts~\cite{karanja2015software}. \\

\noindent{\em RQ1: How do software engineers in East Africa and Northwestern Europe differ in coding skills and problem-solving strategies?}
We approached this question with quantitative methods, including standardized coding and system design tests that measure programming skills and solution complexity. We analyzed the results using machine learning and clustering techniques for comparative insights. In addition, we conducted in-depth on-site interviews to explore the reasoning behind the strategies.\\

\noindent{\em RQ2: How do communication styles and organizational practices differ between regions?}
We explored this question by applying sentiment analysis to code comments and expressions used when solving problems. Quantitative data highlight differences in process steps across countries, while qualitative interviews with team leads provide insights into organizational cultures.\\

\noindent{\em RQ3:  What perception biases exist in software development across these regions, and how might they influence collaboration between engineers?}
We investigated this question through a perception study that quantitatively assesses biases and through conversations with engineers about their responses. At later stages, we asked interviewees to reflect on the results of the perception study.
\vspace{-1em}
\subsection{Data Collection}
This study draws on data from 48 software engineers, with 10–15 participants each from Rwanda, Uganda, Sweden, and the Netherlands. Participants represented different levels of seniority, with {\em juniors} defined as having under three years of experience and {\em seniors} as having over three years, to capture variation in experience and perspective. Each session was conducted on-site in Rwanda, Uganda, Sweden, and the Netherlands, lasting 60–120 minutes and covering coding, system design, and code review tasks, followed by an in-depth interview. To ensure comparability, participants could use a programming environment of their choice, but they were not permitted to access the internet or AI tools.

The sample size was carefully chosen to ensure validity. Prior research shows that data saturation in qualitative studies is typically reached after 9–17 interviews within homogeneous groups~\cite{hennink2022sample, dagan2023building}, and that most themes are identified within the first dozen interviews~\cite{guest2006many}. With 10–15 participants per country, this study ensures saturation at the national level, while 48 interviews in total provide breadth for cross-cultural comparison. Recruitment relied on professional networks through convenience sampling~\cite{sedgwick2013convenience}, a pragmatic choice in exploratory cross-cultural research where access can be limited. This approach supports the study’s goal of capturing authentic perspectives from practicing engineers across regions. Software engineers share core training and practices, providing a relatively homogeneous group, while the inclusion of four countries adds the heterogeneity needed for comparative analysis.

To identify comparative patterns of skills, practices, and perceptions across regions, we employed a mixed-methods design. Quantitative analyses highlight broad trends, while qualitative interviews provide contextual depth for interpreting those patterns~\cite{karasz2009qualitative, creswell2011best}. Prior work emphasizes that even when quantitative precision is limited, such indicators remain valuable for triangulation when complemented with rich qualitative insights~\cite{hespanhol2019understanding}. To strengthen validity, all tasks and interview questions were grounded in real-world engineering practices. They were designed using competencies from the scientific literature and refined with feedback from experienced practitioners. 

The tasks also covered six of the ten SWEBOK knowledge areas, including software design, construction, testing, quality, process, and maintenance, ensuring coverage of core engineering skills~\cite{colomo2018niveles}. Reliability was supported by administering identical tasks under the same conditions to all participants, applying predefined scoring criteria, and conducting pilot testing to confirm clarity. To safeguard integrity, coding and design tasks were monitored to ensure independent work, and in virtual interviews, a second observer was present to validate the process.

\subsection{Assessments}\label{assessment}
We designed three tasks to evaluate both technical and communication skills, complemented by a study of development process estimates. Together, these assessments targeted three dimensions: technical skills, communication styles, and perceptions.  \\

\noindent{\em Task1:  Programming Assignment (Unique Pairs to Target).} To evaluate algorithmic thinking and coding efficiency, participants solved a variation of the well-known two-sum problem \cite{leetcode_two_sum}. Given a list of integers and a target value, they were asked to return all unique pairs of numbers that summed to the target. For example:
\begin{small}
\begin{verbatim}
Input: nums = [1, 5, 3, 7, 4], target = 8 
Output: [(1, 7), (5, 3)]
Input: nums = [2, 2, 4, 4], target = 8 
Output: [(4, 4)]
\end{verbatim}
\end{small}

The task is simple enough to complete within a short time yet complex enough to differentiate skill levels. Participants typically used one of three strategies: (1) the \textit{brute-force} approach that checked all possible pairs in $O(n^2)$ time, which is easy to implement but inefficient; (2) the \textit{sorting and two-pointer technique} that first sorted the list and then used two indices moving from opposite ends to find pairs in $O(n \log n)$ time; and (3) the most efficient solution, the \textit{hash set method}, that tracked seen numbers and identified complements in a single pass in $O(n)$ time.

To analyze submissions, we used CodeBERT \cite{feng2020codebert} to generate embeddings of the code, reduced them with PCA~\cite{abdi2010principal}, and applied K-means clustering~\cite{sinaga2020unsupervised} to identify structural patterns. This automated analysis was complemented by manual inspection, which classified whether participants used brute force, two-pointer, or hash set solutions, and noted their reasoning during interviews. Together, these analyses provided both quantitative and qualitative insights into regional problem-solving strategies.\\

\noindent{\em Task 2: System Design Challenge (URL Shortener).} To evaluate architectural design, scalability, and data handling, participants completed a system design challenge to outline a URL shortening service similar to {\tt Bit.ly}. The task required them to ensure efficient request handling, unique short links, reliable storage and retrieval, and accurate redirection. Participants explained their solutions verbally in real time, which enabled the assessment of both technical reasoning and communication clarity.  

This challenge reflects practical, real-world scenarios and provides a fair basis for comparing engineers across regions. Prior studies show that software engineers often lack practical design skills, particularly in scalability and architectural reasoning~\cite{wan2023software}. This task addresses this skills gap by examining whether participants could design a scalable and maintainable system. The results were aggregated manually by reviewing each answer and identifying overall country-level patterns, which are presented in Section~\ref{ch:results}.\\

\noindent{\em Task 3: Code Review Challenge (Authentication System).}
To evaluate both security awareness and feedback practices, participants reviewed a deliberately flawed authentication system and proposed improvements via code comments. This task assessed vulnerability detection and the clarity/constructiveness of review feedback, enabling cross-regional comparison. Code reviews improve software quality~\cite{Bacchelli2013ModernCR}, and including security defects reflects expectations that developers contribute to secure software even when not security specialists~\cite{assal2025software}. Listing~\ref{lst:auth_system} shows the code provided to participants for review.
\begin{lstlisting}[
    style=pythonstyle,
    caption={Insecure authentication system provided to participants for the code review challenge.},
    label={lst:auth_system}
]
# This authentication system provides basic login and
# password management functionality

class AuthSystem:
    def __init__(self):
        # Hardcoded credentials and plaintext password
        # storage (insecure)
        self.users = {"admin": "password123"}
        # TODO: replace with secure storage

    def login(self, username, password):
        # Unsafe equality instead of hashed comparison
        if (username in self.users and
                self.users[username] == password):
            print("Login successful!")
            return True

        # No rate limiting or account lockout
        return False

    def change_password(self, username, new_password):
        # No authentication or authorization check
        # No strength validation, hashing, or audit log
        self.users[username] = new_password

# Example usage
auth = AuthSystem()
auth.login("admin", "password123")
auth.change_password("admin", "newpass")
auth.login("admin", "newpass")
\end{lstlisting}

Participants identified common flaws, including plaintext password storage, hardcoded credentials, no authentication for password changes, weak error handling, missing validation, and a lack of rate limiting. We categorized 11 vulnerability issues in total and compared how frequently juniors and seniors in each region identified them. The participant comments referring to vulnerability categories are classified as {\em positive} (supportive and constructive), {\em neutral} (descriptive and technical), and {\em negative} (direct and critical). Some examples of these comments are shown below:
\begin{itemize}
  \item Positive comment: \texttt{Good starting point; consider hashing passwords (e.g., bcrypt) and adding rate limiting.}
  \item Neutral comment: \texttt{Passwords are compared directly; no hashing or authentication on change\_password().}
  \item Negative comment: \texttt{Storing plaintext passwords is unacceptable; remove hardcoded creds and implement hashing.}
\end{itemize}

We further analyzed feedback comments with VADER \cite{hutto2014vader, obaidi2021development} to compare feedback styles. For each participant's comment set, VADER outputs four scores: {\em negative}, {\em neutral}, {\em positive}, and {\em compound} scores. Negative sentiment score was the proportion of text expressing negative emotions; neutral sentiment score was the proportion of text expressing neutral emotions; positive sentiment score was the proportion of text expressing positive emotions; and compound score was the normalized, weighted composite score between -1 and 1~\cite{hutto2014vader}. We calculated the average sentiment score for each country and then (1) cataloged which security issues each participant identified, (2) summarized detection rates by region and seniority, and (3) compared comment tone via the per-engineer VADER scores above.\\

\noindent{\em Development Process Estimation.}
In addition to technical assessments, participants estimated the time required for common activities in the software lifecycle. These activities were grouped into four categories, reflecting prior work on developer workflows~\cite{meyer2019today, licorish2017exploring}:

\begin{itemize}
  \item {\em Issue lifecycle}: begins with investigation and root cause analysis, continues with implementation planning and design, followed by development and testing, then documentation and pull request preparation, and concludes with issue verification and closure.  

  \item {\em Code review process}: begins with code cleanup and documentation, continues with running tests locally and peer review, then addressing feedback, and concludes with final approval.  

  \item {\em Deployment process}: begins with build and test automation, continues with staging deployment and testing, then stakeholder review, and concludes with production deployment and post-deployment verification.  

  \item {\em Bug fix process}: begins with bug investigation and reproduction, continues with fix implementation and testing, then documentation and review, and concludes with deployment and verification.  
\end{itemize}

To analyze these responses, all time estimates were normalized into minutes, ensuring comparability across formats (minutes, hours, days, weeks). We then constructed process flow visualizations, highlighting both the relative time spent per activity and the sequence of tasks. Further analyses identified the top and bottom steps by time consumption, as well as critical path bottlenecks consuming more than 10\% of total effort. Cross-country comparisons were finally aggregated into tables, providing insights into where East African and Northwestern European software engineers focus their efforts and how development practices diverge \cite{chen2015may}. We present these results in Section~\ref{ch:results}.

%% file: sections/4_results.tex
\section{Results and Analysis}\label{ch:results}
This section presents the findings of the comparative study of software engineers in East Africa (Rwanda and Uganda) and Northwestern Europe (Sweden and the Netherlands). Results are organized by research question we discussed in Section~\ref{sec:rq}: {\em RQ1: technical skills and problem-solving}, {\em RQ2: communication and process practices}, and {\em RQ3: biases and perceptions}. All quantitative analyses are complemented with qualitative interview insights.  

\subsection{Technical Skills and Problem-Solving (RQ1)}
In answer to RQ1, we examine the three tasks introduced in Section~\ref{assessment}. Below we present the results of each task. \\

\noindent{\em Task 1: Programming Assignment.} Table~\ref{tab:approach_distribution} summarizes the solution strategies. Engineers from Sweden and the Netherlands predominantly used efficient hash-based approaches, with some variation toward two-pointer techniques. Rwandan participants mostly relied on nested loops, though a minority also used hash maps. Interviews revealed that several Rwandan participants explicitly recognized more efficient methods, even if they did not implement them. Ugandan engineers showed the highest failure rate, with over half of their submissions nullified.  

\begin{table}[h]
\centering
\caption{Distribution of solution approaches by country}
\vspace{-1em}
\footnotesize
\begin{tabular}{|l|c|c|c|c|}
\hline
\textbf{Country} &
\makecell{\textbf{Hash}\\\textbf{map/Set}} &
\makecell{\textbf{Nested}\\\textbf{For Loops}} &
\makecell{\textbf{Two-}\\\textbf{Pointer/Counter}} &
\makecell{\textbf{Failed/}\\\textbf{Other}} \\
\hline
Rwanda & 23\% & 69\% & 0\% & 8\% \\
\hline
Uganda & 29\% & 14\% & 0\% & 57\% \\
\hline
Sweden & 80\% & 10\% & 10\% & 0\% \\
\hline
Netherlands & 55\% & 27\% & 18\% & 0\% \\
\hline
\end{tabular}
\label{tab:approach_distribution}
\end{table}

\begin{figure}[t]
    \centering
    \begin{minipage}{0.55\textwidth}
        \centering
        \includegraphics[width=0.8\textwidth]{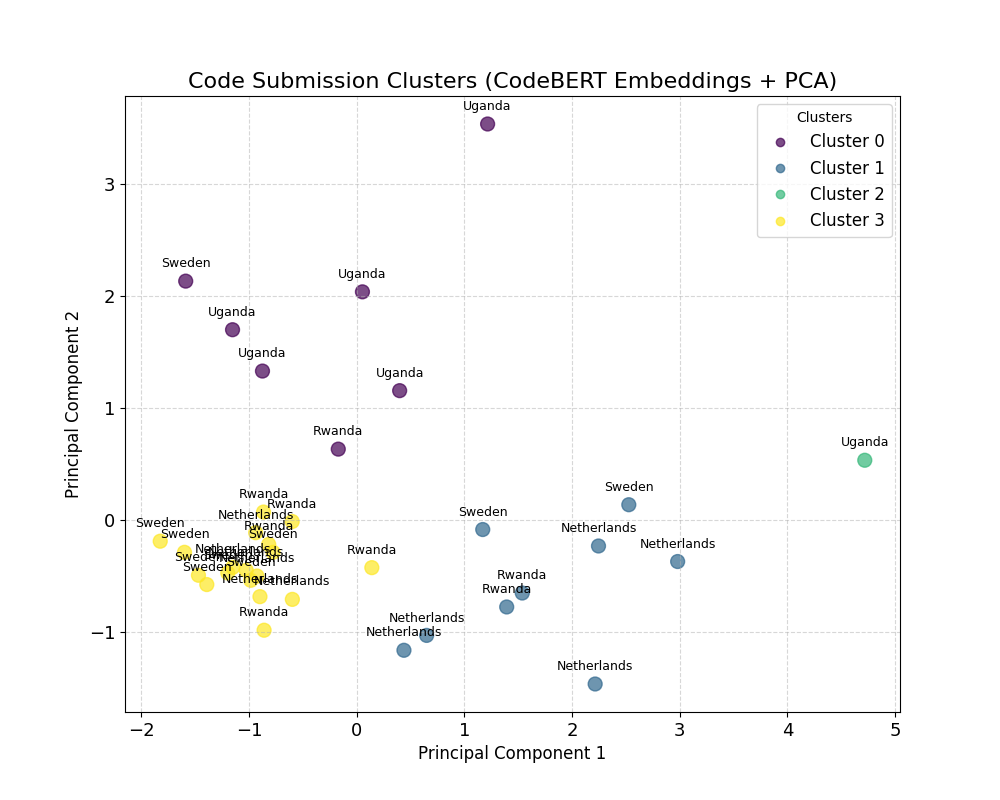}
        \caption{Code submission clusters using CodeBERT embeddings projected via PCA.}
        \label{fig:codebert_clusters}
    \end{minipage}\hfill
    \begin{minipage}{0.4\textwidth}
        \centering
        \includegraphics[width=\textwidth]{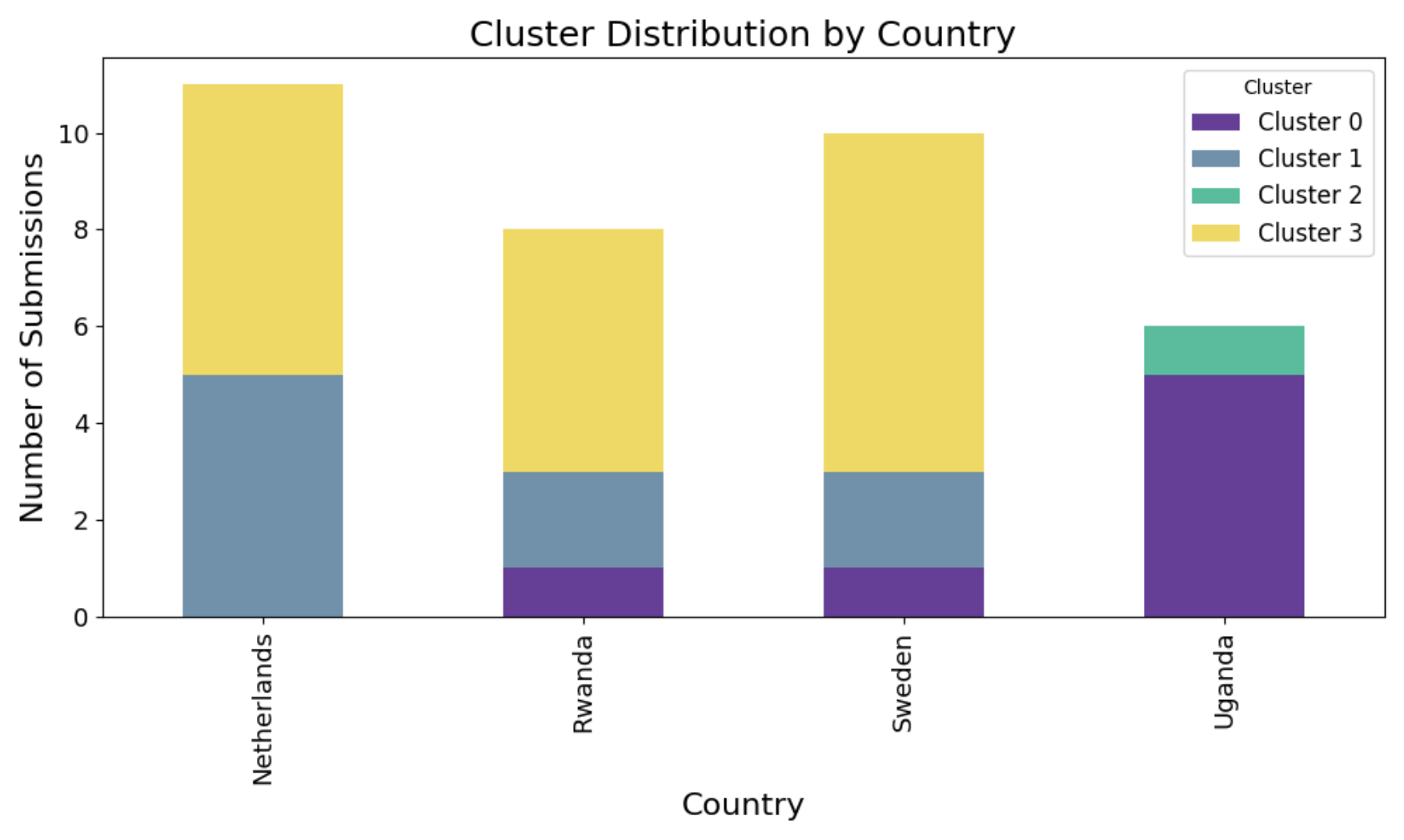}
        \caption{Code submission clusters using CodeBERT embeddings shown through a bar chart.}
        \label{fig:codebert_bargraph}
    \end{minipage}
\end{figure}

To further examine coding styles, we used CodeBERT~\cite{feng2020codebert} embeddings projected with PCA~\cite{abdi2010principal} and clustered with K-means~\cite{sinaga2020unsupervised} (we considered four clusters, representing four different coding styles). Figures~\ref{fig:codebert_clusters} and~\ref{fig:codebert_bargraph} show clear groupings: engineers from Sweden and the Netherlands cluster tightly, reflecting consistent coding conventions, while Rwandan engineers overlap substantially with Europeans despite their reliance on nested loops. Ugandan valid submissions diverged, concentrated in alternative clusters. The clusters can be represented in another way using a barchart as shown in Figure \ref{fig:codebert_bargraph}. Interviews suggested that European consistency reflects standardized curricula, whereas Rwandan variability stems from heterogeneous educational backgrounds and uneven training opportunities. \\

\noindent{\em Task 2: System Design Challenge.}  
The system design task revealed marked regional differences in architectural maturity and variation across experience levels. Engineers from the Netherlands consistently delivered systematic and well-rounded designs, emphasizing security, monitoring, and DevOps practices. Their responses showed little difference between juniors and seniors, reflecting strong knowledge transfer within teams. Swedish engineers demonstrated clear progression by experience: juniors provided solid foundational designs, while seniors expanded solutions with caching, performance optimization, and service separation.  

Rwandan engineers displayed the widest variation. Some gave incomplete or basic answers, while others produced advanced solutions rivaling European seniors, including serverless and document-oriented architectures. Variation was evident across both juniors and seniors, which interviews linked to uneven access to education and professional opportunities. Nonetheless, standout Rwandan engineers showed strong individual potential. Overall, Dutch engineers were the most consistent, Swedish engineers showed the clearest progression with experience, and Rwandan engineers highlighted both challenges and promise. Ugandan submissions were excluded due to integrity concerns and the use of AI tools.  \\

\noindent{\em Task 3: Code Review Challenge.}  
Analyzing Task 3 is divided into two parts: Part 1 focuses on security knowledge, and Part 2 involves code review. Here we focus on Part 1, which directly addresses RQ1, and we will discuss Part 2 under RQ2. Security issue identification rates varied considerably across regions and seniority levels. Engineers from Sweden and the Netherlands consistently detected common vulnerabilities such as plaintext password storage and hardcoded credentials. Rwandan seniors, however, outperformed their European peers in breadth of issues identified, while Rwandan juniors were the weakest group overall. Swedish juniors led in awareness, followed by Dutch, with Rwandan juniors trailing. These findings point to a wide senior–junior gap in Rwanda, where seniors rivaled or exceeded European peers, but juniors performed below all groups. For this task, Ugandan submissions were again excluded due to integrity concerns and the use of AI tools.

\begin{figure}[h]
    \centering
    \includegraphics[width=0.45\textwidth]{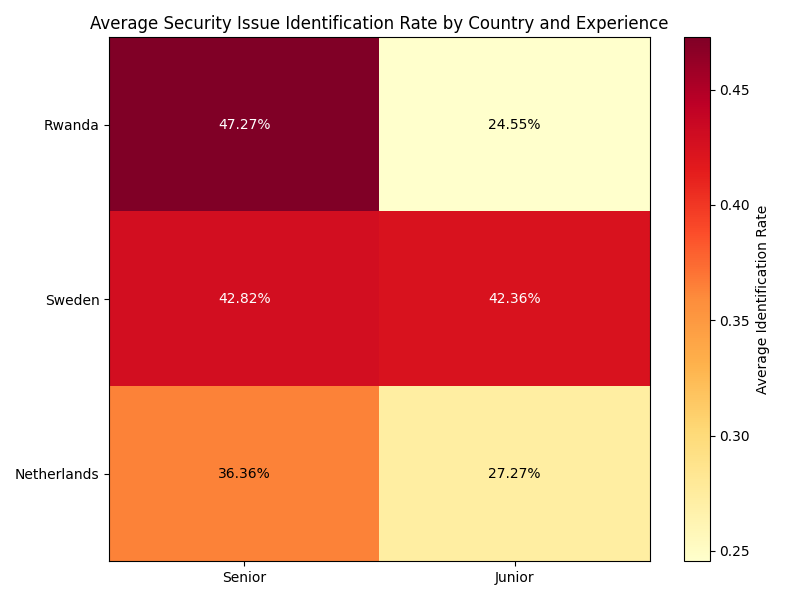}
    \caption{Heatmap comparison of compounded security awareness between senior and junior software engineers across Rwanda, Sweden, and the Netherlands.}
    \label{fig:senjun-heatmap}
\end{figure}

Figure~\ref{fig:senjun-heatmap} highlights that across all countries, seniors outperformed juniors, but Rwandan seniors stood out, demonstrating broader coverage of vulnerabilities than Dutch seniors and nearly matching Swedish peers. Interviews suggested that heavier workloads and greater exposure to diverse tasks in Rwanda may accelerate senior competence, while limited access to quality education explains junior underperformance. Rwandan seniors often accumulate more tech-work hours, with one remarking that in Rwanda ``work is not only work, but survival'' unlike in Europe. Overall, the results show that while European engineers employ more standardized approaches, Rwandan engineers display higher variability: some produced outstanding performances comparable to European standards, while others failed to implement complete solutions. This variability reflects systemic inequalities in education and training opportunities, yet the standout performances illustrate substantial latent potential within East Africa's software engineering workforce.

\subsection{Communication and Process Practices (RQ2)}
To address this research question, we examined two complementary activities: (1) the sentiment expressed in code review comments (Part 2 of Task 3), which reflects communication styles and feedback practices, and (2) self-reported process time estimates, which provide insight into how engineers allocate effort across different stages of the development lifecycle. \\

\noindent{\em Code Review Sentiment.}
The code review challenge examined the sentiment of developer comments, with results summarized in Table~\ref{tab:sentiment_analysis}. While neutrality dominated across all countries, the compound sentiment scores highlight clear contrasts. Rwandan and Ugandan engineers provided more positive and supportive feedback (compound scores $>$ 0.5, close to +1), suggesting an emphasis on encouragement. In contrast, Dutch engineers leaned toward more negative tones, while Swedish engineers remained largely neutral.  

Interviews revealed that European engineers generally preferred direct critique, which they viewed as constructive and aligned with common norms in their engineering cultures. East African engineers, by contrast, valued supportive tones that reinforced collaborative spirit. These cultural differences have important implications for cross-cultural collaboration: while Rwandan and Ugandan teams may foster more emotionally supportive environments, Swedish and Dutch teams may favor task-oriented directness.\\
\begin{table}[t]
\centering
\caption{Sentiment analysis of code review comments.}
\footnotesize
\begin{tabular}{|l|c|c|c|c|}
\hline
\textbf{Country} & Negative & Neutral & Positive & Compound \\
\hline
Rwanda & 0.049 & 0.818 & 0.133 & +0.541 \\
\hline
Sweden & 0.113 & 0.780 & 0.107 & +0.007 \\
\hline
Netherlands & 0.153 & 0.764 & 0.083 & -0.361 \\
\hline
Uganda & 0.047 & 0.713 & 0.240 & +0.919 \\
\hline
\end{tabular}
\label{tab:sentiment_analysis}
\vspace{-2em}
\end{table}

\noindent{\em Process Time Estimates.}
Table~\ref{tab:process-times} shows the self-reported software development process time estimates by country. This table is divided into four process steps, where each is broken down into finer-grained tasks. Participants were asked to estimate how long each task typically takes them to complete. Table~\ref{tab:critical_path} shows the steps that take $>10\%$ of the total time. The results reveal clear regional emphases. Swedish and Dutch engineers devoted most of their time to early development stages such as planning, design, and implementation, while Rwandan engineers emphasized deployment activities and post-deployment verification. Ugandan engineers reported much longer durations across nearly all steps. Interviews suggested these inflated times stemmed largely from English-as-a-second-language challenges rather than genuine inefficiencies. For this reason, Uganda is included in the table for transparency but treated as an outlier in cross-country comparisons.

Comparisons of the fastest and slowest tasks provide further insights. Rwandan engineers were relatively quick at bug investigation, documentation, and deployment, but slower in development and testing. Dutch engineers minimized time on deployment checks and approvals but faced bottlenecks in planning and bug fixing. Swedish engineers completed code review tasks quickly but spent considerable time on build and test automation, planning, and development. Critical path analysis (Table~\ref{tab:critical_path}) reinforces these contrasts: engineers from Sweden and the Netherlands concentrated their effort on development and testing, while Rwandans invested a larger share in deployment, including automation and post-deployment verification.

Overall, the findings suggest distinct optimization strategies. European engineers seek to reduce rework and technical debt through heavy upfront investment in planning and development, whereas Rwandan engineers focus on robustness by emphasizing collaborative review, deployment, and verification. These differences reflect broader organizational and cultural approaches to balancing efficiency with reliability in software development.

\begin{table*}[t] 
\centering 
\caption{Self-reported software development process time (minutes) estimates by country.} 
\vspace{-1em}
\footnotesize 
\begin{tabular}{|p{2.5cm}|p{3.5cm}|c|c|c|c|} 
\hline 
\textbf{Process} & \textbf{Step} & \textbf{RW} & \textbf{NL} & \textbf{SE} & \textbf{UG} \\ 
\hline 
Issue & Development/testing & 261.1 & 720.0 & 792.0 & 1920.0 \\ 
Lifecycle & Documentation/PR & 101.1 & 96.0 & 109.0 & 960.0 \\ 
& Planning/design & 194.4 & 189.0 & 264.0 & 1600.0 \\ 
& Investigation/analysis & 165.9 & 111.0 & 132.0 & 1240.0 \\ 
& Verification/closure & 141.7 & 51.0 & 88.0 & 840.0 \\ 
\hline 
Code Review & Addressing feedback & 143.9 & 64.5 & 90.0 & 1380.0 \\ 
Process & Cleanup/documentation & 236.1 & 58.5 & 76.5 & 1140.0 \\ 
& Final approval & 82.2 & 34.5 & 51.0 & 1190.0 \\ 
& Peer review process & 87.2 & 96.0 & 93.0 & 1360.0 \\ 
& Running tests locally & 132.4 & 57.0 & 57.5 & 1170.0 \\ 
\hline Deployment & Build/test automation & 287.8 & 135.0 & 264.0 & 1145.0 \\ 
Process & Post-deploy verification & 335.6 & 51.0 & 58.0 & 1400.0 \\ 
& Production deployment & 126.7 & 54.0 & 72.0 & 1120.0 \\ 
& Staging deploy/testing & 12.0 & 64.5 & 70.0 & 1080.0 \\ 
& Stakeholder review & 124.4 & 61.5 & 76.5 & 1360.0 \\ 
\hline 
Bug Fix & Investigation/reproduction & 60.0 & 147.0 & 213.0 & 1400.0 \\ 
Process & Deployment verification & 45.6 & 78.0 & 96.0 & 980.0 \\ 
& Documentation/review & 25.6 & 52.5 & 54.0 & 1080.0 \\ 
& Implementation/testing & 46.1 & 261.0 & 261.0 & 1120.0 \\ 
\hline \multicolumn{6}{p{11cm}}{\small Note: Values represent average time in minutes. RW=Rwanda, NL=Netherlands, SE=Sweden, UG=Uganda.} 
\end{tabular} 
\label{tab:process-times} 
\end{table*}

\begin{table}[htbp] 
\centering 
\caption{Critical path analysis: steps taking >10\% of total time in minutes} 
\vspace{-1em}
\footnotesize 
\begin{tabular}{|c|l|r|r|} 
\hline 
\textbf{Country} & \textbf{Process Step} & \textbf{Time} & \textbf{\%} \\
\hline
RW & Dev \& testing (Issue Lifecycle) & 261.1 & 10.0\% \\
& Post-deploy verification (Deployment) & 335.6 & 12.9\% \\
& Build \& test automation (Deployment) & 287.8 & 11.0\% \\
\hline
NL & Dev \& testing (Issue Lifecycle) & 720.0 & 30.2\% \\
& Fix impl \& testing (Bug Fix) & 261.0 & 11.0\% \\
\hline
SWE & Dev \& testing (Issue Lifecycle) & 792.0 & 27.1\% \\
\hline
\end{tabular}
\label{tab:critical_path} 
\vspace{-1em}
\end{table}
 
\subsection{Biases and Perceptions (RQ3)}
Unlike RQ1 and RQ2, which examined measured skills and practices, RQ3 focuses on how software engineers perceive their own and others’ competencies. To this end, we analyze the ratings given by participants from each country to all four countries across three skill areas: (1) software development, (2) problem solving, and (3) communication. The perception ratings revealed consistent regional biases.

\begin{figure*}[t]
    \centering
    \begin{minipage}{0.45\textwidth}
        \centering
        \includegraphics[width=0.9\textwidth]{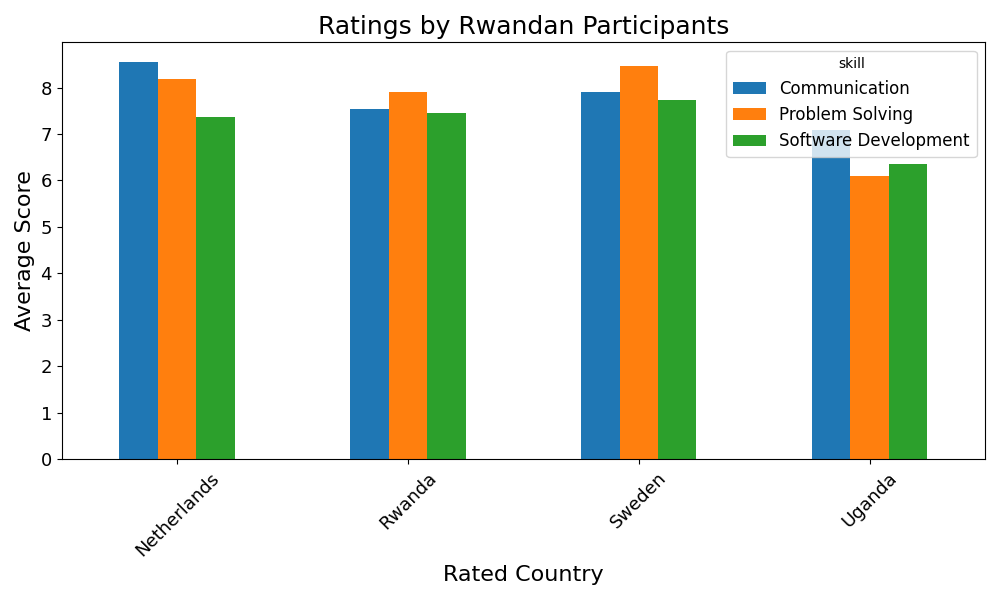}
        \caption{\small Perception-based given by Rwandan participants to developers from Rwanda, Sweden, the Netherlands, and Uganda.}
    \label{fig:rwanda-bias}
    \end{minipage}\hfill
    \begin{minipage}{0.45\textwidth}
        \centering
        \includegraphics[width=0.9\textwidth]{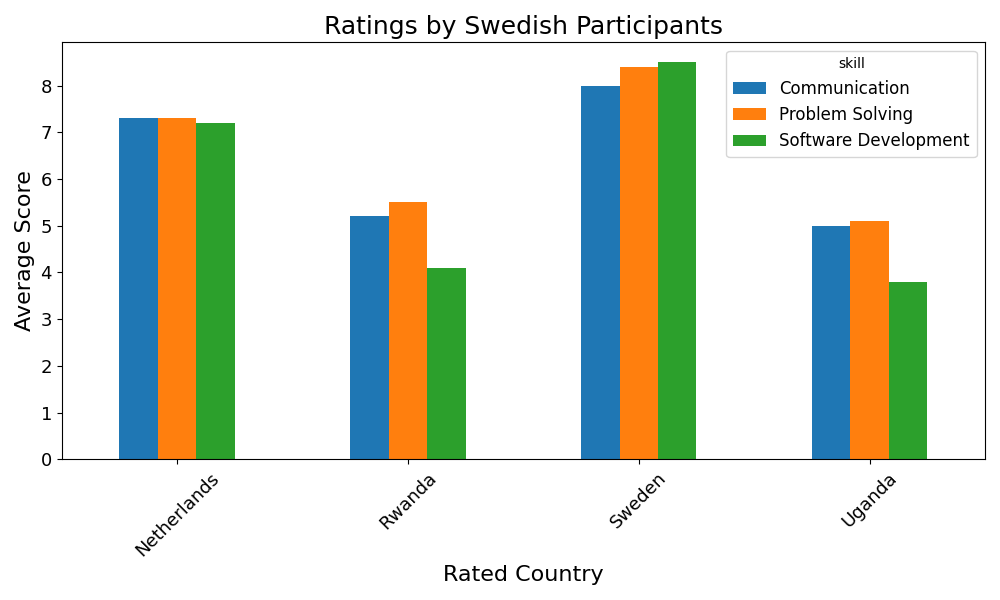}
        \caption{\small Perception-based ratings given by Swedish participants to software engineers from Rwanda, Sweden, the Netherlands, and Uganda.}
        \label{fig:sweden-bias}
    \end{minipage}

    \vspace{0.5em}

    \begin{minipage}{0.45\textwidth}
        \centering
        \includegraphics[width=0.9\textwidth]{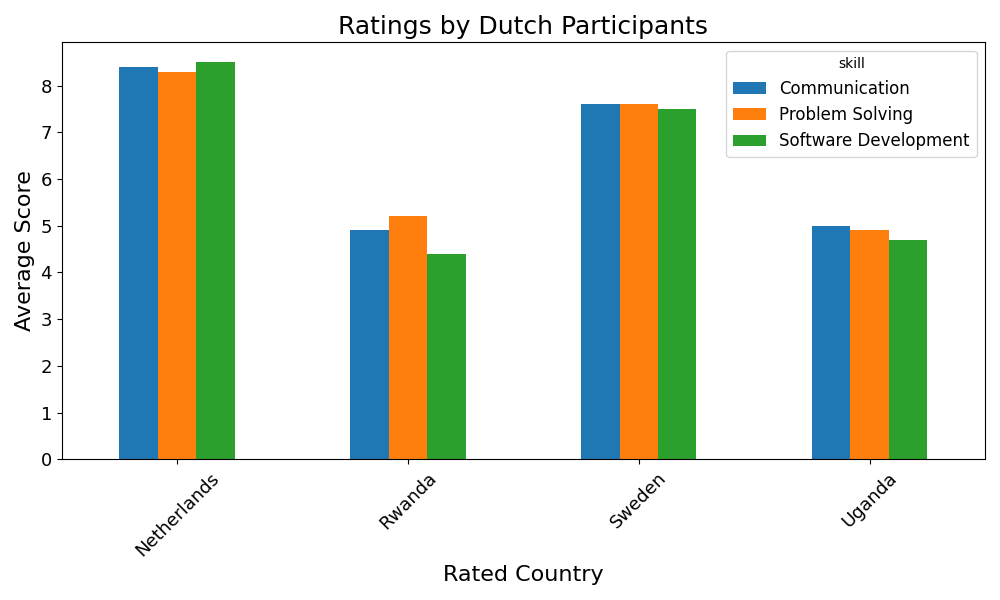}
        \caption{\small Perception-based ratings given by Dutch participants to software engineers from Rwanda, Sweden, the Netherlands, and Uganda.}
        \label{fig:netherlands-bias}
    \end{minipage}\hfill
    \begin{minipage}{0.45\textwidth}
        \centering
        \includegraphics[width=0.9\textwidth]{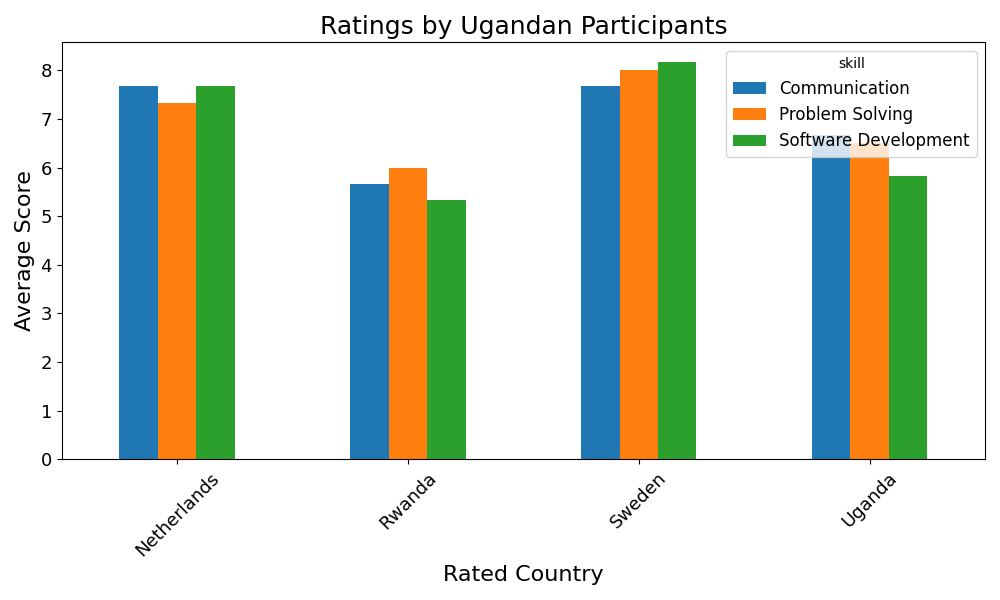}
        \caption{\small Perception-based ratings given by Ugandan participants to software engineers from Rwanda, Sweden, the Netherlands, and Uganda.}
        \label{fig:uganda-bias}
    \end{minipage}
\end{figure*}

Rwandan participants generally saw themselves as comparable to Europeans, though they rated Ugandan engineers slightly lower (Figure~\ref{fig:rwanda-bias}). Many emphasized that Ugandan developers are capable, yet still ranked them beneath Europeans and themselves. They also noted differences between Europeans, describing engineers from Sweden as entrepreneurial and Dutch engineers as hardworking.

Swedish engineers rated themselves the highest, followed by Dutch peers, while placing both Rwandan and Ugandan engineers much lower (Figure~\ref{fig:sweden-bias}). They attributed this not to individual ability but to systemic issues such as weaker education and infrastructure in East Africa. Dutch engineers ranked themselves slightly above Swedes and rated East African engineers as much less capable (Figure~\ref{fig:netherlands-bias}). Interviews revealed that many Dutch participants lacked detailed knowledge about Rwanda or Uganda, often generalizing based on stereotypes of ``Africa'' rather than direct experience. Ugandan engineers rated themselves slightly higher than Rwandans but still below Europeans (Figure~\ref{fig:uganda-bias}), citing superior education and infrastructure in Northwestern Europe. This internalized hierarchy underscores how systemic disparities shape professional confidence.

On average, across all groups, European engineers consistently rated themselves highest and African engineers lowest. Interviews confirmed that participants expected such results: Rwandan team leads anticipated bias against them, while European team leads saw the ratings as reflecting “real” skill distributions. Yet, performance data contradicts these assumptions, as Rwandan seniors often matched or exceeded European peers in technical tasks. These findings reveal a gap between measured ability and perceived competence. While East African engineers expressed relatively balanced views of global capabilities, European engineers systematically underrated their African counterparts. Such perception biases risk undermining equitable collaboration by obscuring complementary strengths and reinforcing global hierarchies within software engineering.

%% file: sections/5_discussion.tex
\section{Discussion}
\label{sec:discussion}
This section discusses the main findings of the comparative study and their implications for global software development by addressing the three research questions introduced in Section~\ref{sec:rq}. Overall, the results indicate that differences between East Africa and Northwestern Europe are shaped less by inherent capability and more by systemic factors, cultural practices, and perception biases.\\

\noindent\textbf{Technical Skills and Prblem-Solving (RQ1)}. The technical assessments revealed consistent performance among developers in Northwestern Europe and higher variance in East Africa, particularly in Rwanda. While many Rwandan juniors underperformed, several seniors matched or surpassed European peers, especially in security awareness. These disparities appear to be systemic: European engineers benefited from standardized education and structured career paths, whereas Rwandan engineers described uneven university training and gaps in algorithms or data structures that persisted into their early careers. At the same time, standout East African engineers clustered closely with Europeans in the CodeBERT analysis, especially those with exposure to international consultancies, open-source contributions, or graduate training. Rwandan seniors also reported much longer working hours (often 60+ per week across multiple projects), which may accelerate experiential learning and explain their advanced security knowledge despite weaker formal education. For organizations, this suggests that East African talent should be recognized for its diversity, where some engineers already perform at global standards and others are still developing, with opportunities and exposure playing a key role in shaping outcomes.\\

\noindent\textbf{Communication and Process Practices (RQ2)}. Communication styles reflected regional cultural patterns. Sentiment analysis of code reviews showed that Rwandan and Ugandan engineers provided more positive and supportive feedback (compound scores $>0.5$), while Dutch engineers leaned toward negative, and Swedish engineers were neutral. Interviews confirmed that Europeans considered direct critique efficient and professional, while East African engineers valued encouragement and group harmony. These align with Hofstede's framework: higher power distance and collectivism in East Africa encourage supportive communication, while lower power distance and individualism in Northwestern Europe favor directness. Process allocations revealed similar contrasts. Dutch and Swedish engineers devoted around 30\% of effort to initial design and development, emphasizing strong architecture to minimize rework. Rwandan engineers focused more on deployment, verification, and review, often citing rigorous DevOps pipelines and collective bug fixing. Rather than inefficiency, this reflects different optimization strategies: Northwestern Europe teams invest upfront, while East African teams distribute risk across collaborative verification. In global teams, these strategies could be complementary, combining robustness in design with resilience in testing and deployment.\\

\noindent\textbf{Biases and Perceptions (RQ3)}. Perception studies revealed gaps between measured skills and perceived ability. European developers consistently rated themselves and one another higher, often scoring East African peers below 5/10, despite evidence of individual excellence among Rwandan seniors. Many admitted to having limited knowledge of East Africa's ecosystems, instead generalizing from stereotypes of weak infrastructure and education. By contrast, Rwandans tended to view themselves as equal to Europeans but superior to Ugandans, while Ugandans rated Europeans highest, reflecting internalized hierarchies shaped by systemic inequities. These perceptions risk undermining confidence and collaboration: Rwandan seniors who matched European performance still expected to be underestimated. Addressing such biases is as crucial for global software development as addressing technical gaps, since they can erode trust and prevent recognition of complementary strengths.\\

\noindent\textbf{Methodological Reflections}.
The pragmatic mixed-methods design enabled triangulation across programming tasks, design challenges, process estimates, and interviews. Strengths include bootstrap resampling that confirmed the stability of clustering results and high inter-rater reliability in perception ratings. However, the sample was small (10–15 per country) and recruited via professional networks, skewing toward high achievers in Europe and Kigali-based consultancies. Ugandan data was further limited by language barriers and unauthorized AI use. Findings should therefore be interpreted as exploratory rather than definitive. Future research should employ larger, possibly longitudinal, samples to track skill evolution over time, incorporate more diverse East Africa contexts beyond capital cities, and complement self-reports with log data for higher precision. Nevertheless, this study shows the value of pragmatic, flexible methodologies in capturing both quantitative patterns and contextual nuance in cross-cultural software engineering.\\

\noindent\textbf{Implications}. Taken together, the findings demonstrate that global collaboration is shaped not only by technical ability but also by systemic inequities, cultural practices, and perception hierarchies. With respect to \textbf{RQ1}, East African engineers represent a high-variance talent pool where standout performers reach global standards despite systemic disadvantages. For \textbf{RQ2}, communication and process practices reflect culturally adaptive optimization strategies that could be complementary in cross-regional teams. Finally, \textbf{RQ3} shows that perception biases, both external stereotypes and internalized hierarchies, pose persistent risks for equitable collaboration. Recognizing and addressing these dynamics is critical for building effective and inclusive global software development teams.

%% file: sections/6_conclusion.tex
\section{Conclusion}
\label{sec:conclusion}
This study compared software engineering practices in East Africa (Rwanda and Uganda) and Northwestern Europe (Sweden and the Netherlands), addressing three research questions on technical competence, communication practices, and perception biases in global software development. The findings show that observed differences across regions are better explained by systemic and cultural factors than by inherent technical ability. European engineers demonstrated strong consistency in coding and system design, whereas Rwandan engineers showed greater performance variation, with several senior engineers performing at levels comparable to or exceeding those of their European peers, particularly in security-related tasks. These results challenge assumptions that outsourcing to Africa entails a quality trade-off and instead point to complementary strengths: Northwestern European engineers tend to emphasize upfront design and standardization, whereas East African engineers place greater emphasis on collaborative verification, rigorous DevOps practices, and supportive communication. Despite these complementary capabilities, significant perception gaps persist. European participants systematically underestimated the competence of their East African counterparts, even when objective measures showed comparable or superior performance. This mismatch between measured performance and perceived competence highlights how everyday evaluation practices can reproduce epistemic unfairness and reinforce existing inequalities in global software labor.

Methodologically, the pragmatic mixed-methods design proved valuable for linking quantitative performance with qualitative cultural insights; however, limitations include a small sample size, selection bias, and inconsistent data quality in Uganda. Future research should expand to larger, more diverse samples, include longitudinal tracking of junior progression, and complement self-reports with repository and process log analysis. Developing sentiment analysis tools tailored for technical communication across cultures would further strengthen validity. Overall, this work provides one of the first empirical baselines of cross-cultural software engineering in East Africa. By documenting both technical and cultural dimensions, it contributes to a more nuanced understanding of global collaboration, highlighting the untapped potential of African software engineers when systemic barriers are addressed. The findings underscore that effective global software development depends not only on technical alignment but also on bridging cultural practices and perception gaps.